\documentclass[aps,prl,twocolumn,showpacs,superscriptaddress]{revtex4-1}

\usepackage{amsfonts}
\usepackage{amsmath}
\usepackage[pdftex]{graphicx}
\usepackage{bm}

\begin{document}

%%%%%%last edited by Makariy February 21,2013 referee reports and TS comments

%%%%%%%%%%%%%%%%%%%%%%%%%%%% TITLE

\title{ Inter-plane resistivity of isovalent doped BaFe$_2$(As$_{1-x}$P$_x$)$_2$}

%%%%%%%%%%%%%%%%%%%%%%%%%%%% AUTHORS

\author{M.~A.~Tanatar}
\email[Corresponding author: ]{tanatar@ameslab.gov}
\affiliation{Ames Laboratory, Ames, Iowa 50011, USA}
\affiliation{Department of Physics and Astronomy, Iowa State University, Ames, Iowa 50011, USA }

\author{K.~Hashimoto}
\email{hashimoto@imr.tohoku.ac.jp}
\affiliation{Department of Physics, Kyoto University, Kyoto 606-8502, Japan}

\author{S.~Kasahara}
\email{kasa@scphys.kyoto-u.ac.jp}
\affiliation{Department of Physics, Kyoto University, Kyoto 606-8502, Japan}

\author{T.~Shibauchi}
\email{shibauchi@scphys.kyoto-u.ac.jp}
\affiliation{Department of Physics, Kyoto University, Kyoto 606-8502, Japan}

\author{Y.~Matsuda}
\email{matsuda@scphys.kyoto-u.ac.jp}
\affiliation{Department of Physics, Kyoto University, Kyoto 606-8502, Japan}

\author{R.~Prozorov}
\email{prozorov@ameslab.gov}
\affiliation{Ames Laboratory, Ames, Iowa 50011, USA}
\affiliation{Department of Physics and Astronomy, Iowa State University, Ames, Iowa 50011, USA }

\date{21 February 2013}

%%%%%%%%%%%%%%%%%%%%%%%%%%%% ABSTRACT

\begin{abstract}

Temperature-dependent inter-plane resistivity, $\rho _c(T)$,  was measured for the iron-based superconductor BaFe$_2$(As$_{1-x}$P$_x$)$_2$ over a broad isoelectron phosphorus substitution range from $x$=0 to $x$=0.60, from non-superconducting parent compound to heavily overdoped superconducting composition with $T_c\approx 10~K$. The features due to structural and magnetic transitions are clearly resolved in $\rho _c(T)$ of the underdoped crystals.  A characteristic maximum in $\rho _c(T)$, found in the parent BaFe$_2$As$_2$ at around 200~K, moves rapidly with phosphorus substitution to high temperatures. At the optimal doping, the inter-plane resistivity shows $T$-linear temperature dependence without any cross-over anomalies, similar to the previously reported in-plane resistivity. This observation is in stark contrast with dissimilar temperature dependences found at optimal doping in electron-doped Ba(Fe$_{1-x}$Co$_x$)$_2$As$_2$. Our finding suggests that despite similar values of the resistivity and its anisotropy, the temperature dependent transport in the normal state is very different in electron and isoelectron doped compounds. Similar temperature dependence of both in-plane and inter-plane resistivities, in which the dominant contributions are coming from different parts of the Fermi surface, suggests that scattering is the same on the whole Fermi surface. Since magnetic fluctuations are expected to be much stronger on the quasi-nested sheets, this observation may point to the importance of the inter-orbital scattering between different sheets. 

\end{abstract}


%Metals, transport processes in, 72.15.-v

%Superconducting materials 74.70.Dd Ternary, quaternary, and multinary compounds (including Chevrel phases, borocarbides, etc.)

%74.25.Jb Electronic structure

\maketitle

%%%%%%%%%%%%%%%%%%%%%%%%%%%% INTRODUCTION

%%%%1) Introduction
%Oxypnictide superconductors, discovery, properties

\section{Introduction}

The parent compound of 122 family of iron-arsenide superconductors, BaFe$_2$As$_2$, crystallizes in a tetragonal symmetry ThRh$_2$Si$_2$ structure. The structure undergoes transformation to an orthorhombic on cooling below $T_S$ with concomitant or subsequent magnetic ordering below $T_N \leq T_S$. Superconductivity is induced on suppression of magnetism/orthorhombicity with maximum $T_c$ observed close to a point where $T_S (x)$ and $T_N(x)$ extrapolate to zero. This proximity to a quantum critical point suggests that superconductivity may be magnetically mediated \cite{Mathur,Monthoux}.

A unique feature of the mechanism suggested for magnetically mediated superconductivity \cite{Mathur} is a prediction of the systematic evolution of the electronic properties of the compounds with control tuning parameter. In particular, at a quantum critical point (QCP) a temperature-dependent electrical resistivity is expected to follow a power law function $\rho - \rho_0 =AT^n$ (here $\rho _0$ is residual resistivity due to scattering on impurities and defects) with the exponent $n<2$, different from expectations of Landau Fermi-liquid theory. Away from QCP this dependence transforms towards usual $T^2$- temperature dependence. In iron pnictides this doping-evolution is revealed in temperature-dependent in-plane transport in electron doped Ba(Fe$_{1-x}$Co$_x$)$_2$As$_2$ (BaCo122 in the following) \cite{NiNiCo,NDL}, in isoelectron substituted BaFe$_2$(As$_{1-x}$P$_x$)$_2$ \cite{Kasahara} (BaP122 in the following) and under pressure \cite{pressure}, with $n$=1 in all three cases. In BaP122 the existence of QCP was suggested by NMR \cite{BaP-QCP} and magnetoquantum oscillation studies \cite{BaP-dHvA,SdH-P,Currington} in the normal state. Effect of the QCP was found in the superconducting state as well, as a peak in the doping-dependent value of $T \to 0$ London penetration depth \cite{BaP-Science}. In addition to the critical evolution of magnetism, NMR studies of BaCo122 found temperature-dependent Knight shift, suggesting the existence of a pseudogap \cite{Co-pseudogap1,Co-pseudogap2}. This temperature-dependent Knight shift correlates with a broad cross-over maximum in the temperature-dependent inter-plane resistivity \cite{pseudogap}, and similar crossovers observed for other transition metal dopings \cite{pseudogap2}. It also correlates with onset of pseudogap behavior in spectroscopic measurements in BaCo122 \cite{SJMoon}.
No temperature dependence of Knight shift is observed in optimally doped BaP122 \cite{BaP-QCP}, however, the optical studies find pseudogap in both Co-doped and P-doped compositions \cite{SJMoon}. In the BaP122 case the onset of pseudogap in spectroscopic $ab$-plane reflectivity measurements correlates with the temperatures of appearance of anomalous nematic response in torque measurements \cite{KasaharaNature} and in-plane resistivity anisotropy \cite{BaPdetwinning}. 

In order to get further insight into the normal state anomalies of iron-pnicte superconductors, in this article we perform detailed study of the temperature-dependent inter-plane resistivity of BaP122 over a broad doping range from parent compound through optimal doping ($x_{opt}=$0.33, $T_{c,opt}$=30~K) to heavily overdoped composition with $x \approx$0.60 ($T_c$=10~K).
We find a rapid rise of the inter-plane resistivity crossover temperature $T_{max}$ in the under-doped regime so that a perfectly $T$-linear temperature-dependent inter-plane resistivity is observed at optimal doping up to temperature as high as 400~K, similar to previous observation of $T$-linear dependence in in-plane transport \cite{Kasahara}. This lack of significant features in either in-plane or inter-plane resistivity makes BaP122 system distinct from both electron-doped BaCo122 (linear $\rho_a(T)$ and cross-over $\rho_c(T)$) and hole doped BaK122 \cite{BaKanisotropy}), and electron- and environmentally doped NaFeAs \cite{NaFeAs1,NaFeAs2}, all with cross-overs in both $\rho_a(T)$ and $\rho_c(T)$.
This difference suggests that three-dimensional character of the Fermi surface and normal state scattering are important in iron pnictides, and these are significantly different for different compounds and dopant species.

\section{Experimental}

Single crystals of BaP122 were grown from stoichiometric mixtures of Ba (flakes), and FeAs, Fe, P or FeP (powders) placed in an alumina crucible, sealed in evacuated quartz tube. It was heated to 1150-1200C, kept for 12 hours and then cooled down slowly to 800C at a rate 1.5C/h. Platelet crystals had typically 0.3 to 0.7 mm$^2$ surface area, their $x$ value was determined using energy dispersive electron probe microanalysis (EDX).

Samples for the study were extensively characterized by polarized optics \cite{domains} and magneto-optic techniques \cite{MO} to look for possible inhomogeneity, as described in detail in Ref.~\onlinecite{SUST}. 
Inter-plane resistivity was measured using a two-probe technique, relying on the negligibly small contact resistance. The details of the measurement procedure were the same as in our previous studies on pure and transition metal doped Ba122 compounds, see Refs.~\onlinecite{anisotropy,anisotropypure,pseudogap,pseudogap2} for details. In brief, samples typically had dimensions 0.5$\times$0.5$\times$(0.02-0.1) mm$^3$ ($a\times b\times c$), all sample dimensions were measured with an accuracy of about 10\%. The top and bottom $ab$-plane surfaces were covered with ultrapure Sn solder, as described in Ref.~\onlinecite{SUST}, forming a capacitor-like structure. Tin-soldering technique produced contact resistance typically in the 10 $\mu \Omega$ range. Four-probe scheme was used down to the sample to measure series-connected sample, $R_s$, and contact, $R_c$ resistance. Taking into account that $R_s \gg R_c$, contact resistance represents a minor correction of the order of 1 to 5\%. This can be directly seen for our samples for temperatures below the superconducting $T_c$, where $R_s =$0 and the measured resistance represents $R_c$ \cite{anisotropy,anisotropypure,SUST,vortex}. 

A tendency of the samples to cleave along $ab$ plane leads to a serious problem in inter-plane resistivity measurements. Even in visually perfect crystals, we frequently encounter partial cracks, leading to current redistribution in sample cross-section, and admixture of in-plane resistivity into measured inter-plane resistivity. To control this problem, we used as thin samples as were available and performed measurements of $\rho_c$ on at least 5 samples of each composition. In all cases we obtained qualitatively similar temperature dependencies of the electrical resistivity, as represented by the ratio of resistivities at room and low temperatures, $\rho _c (0)/\rho _c (300)$. The resistivity value, however, showed a notable scatting and at room temperature was typically in the range 1000 to 2000 $\mu \Omega$~cm, which is very similar to all transition metal doped Ba122 \cite{pseudogap,pseudogap2}, as well as for hole-doped BaK122 \cite{BaKanisotropy}.

\section{Results}

\begin{figure}[tbh]%
\centering
\includegraphics[width=8cm]{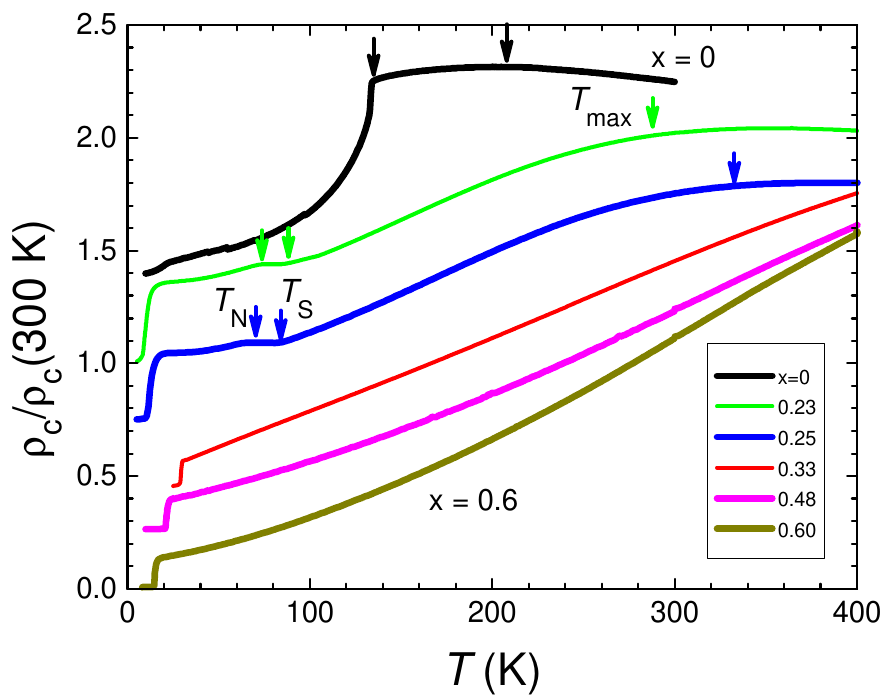}%
\caption{(Color online) Temperature-dependent inter-plane resistivity of BaFe$_2$(As$_{1-x}$P$_x$)$_2$ for (top to bottom) $x$=0 (parent compound, black curve), underdoped compositions $x$=0.23 (green) and $x$=0.25 (blue), optimally doped $x$=0.33 (red) and overdoped $x$=0.48 (magenta) and $x$=0.60 (dark yellow). The data are plotted on the normalized resistivity scale, $\rho_c(T)/\rho_c(300K)$, and offset progressively downwards for higher $x$ to avoid overlapping. Arrows show a position of the resistivity cross-over temperature $T_{max}$ and of the structural, $T_S$, and magnetic $T_N$ transitions. }
\label{interplane}%
\end{figure}

Fig.~\ref{interplane} shows the main experimental result of this paper, a temperature-dependent inter-plane resistivity of BaP122 for several compositions from non-superconducting parent compound, $x$=0, through optimally doped, $x$=0.33 and $T_c$=30~K, to heavily overdoped, $x$=0.60 and $T_c$=10~K . For the sake of comparison the data are plotted on a normalized resistivity scale, $\rho_c(T)/\rho_c(300K)$, and offset downwards for increasing $x$.

%%%%%%Figure1 doping evolution of inter-plane resistivity of BaP122

Several features should be noticed. First, the curves for samples with $x$=0.23 and $x$=0.25 show a clear upward turn on cooling through the temperature of structural transition $T_{S}$ and downturn below the temperature of magnetic transition $T_{N}$, marked with arrows in Fig.~\ref{interplane}. The values of $T_{S}$ and $T_{N}$ are in good agreement with NMR results \cite{BaP-QCP}. This splitting of structural and magnetic transitions in BaP122 is similar to electron-doped BaCo122 \cite{Pratt}. 

An additional feature is clearly observed in $\rho_c(T)$ in parent and underdoped compositions $x$=0.23 and $x$=0.25 at temperatures above 200~K. The $\rho_c(T)$ changes slope and shows a downturn on warming with resistivity taking very shallow maximum at a temperature $T_{max}$ as indicated with arrows.  By comparison with NMR studies in BaCo122 \cite{pseudogap}, and with $\rho _c(T)$ for other transition metal substitutions \cite{pseudogap2}, we previously assigned this maximum in transition metal doped Ba122 to the onset of carrier activation over the pseudogap. Similar assignment was suggested for the explanation of a maximum in $\rho_c (T)$ and a slope-saturation in $\rho _a(T)$ in hole-doped BaK122 and in NaFeAs \cite{BaKanisotropy,NaFeAs1,NaFeAs2}.
Alternatively, the slope change in $\rho _a(T)$ of optimally doped BaK122 was explained in multi-band scenario \cite{zverev}. 

%{A minimum requirement of this model is the presence of two bands with strongly different temperature-dependent conductivities. Conductivity in one band, $\sigma _1$ is nearly independent of temperature, while in another band, $\sigma _2$ it is strongly temperature dependent. At low temperatures $\sigma_2 \gg \sigma _1$ and conductivity is strongly $T$-dependent, at high temperatures $\sigma _2 \ll \sigma_1$ and conductivity saturates. Though very attractive, this model neither explains $\rho _c (T)$ decrease on warming paralleling $\rho _a (T)$ slope change in BaK122 \cite{BaKanisotropy}, nor doping independence of $\rho _c(T)$ maximum position in BaK122. }

\begin{figure}[tbh]%
\centering
\includegraphics[width=8cm]{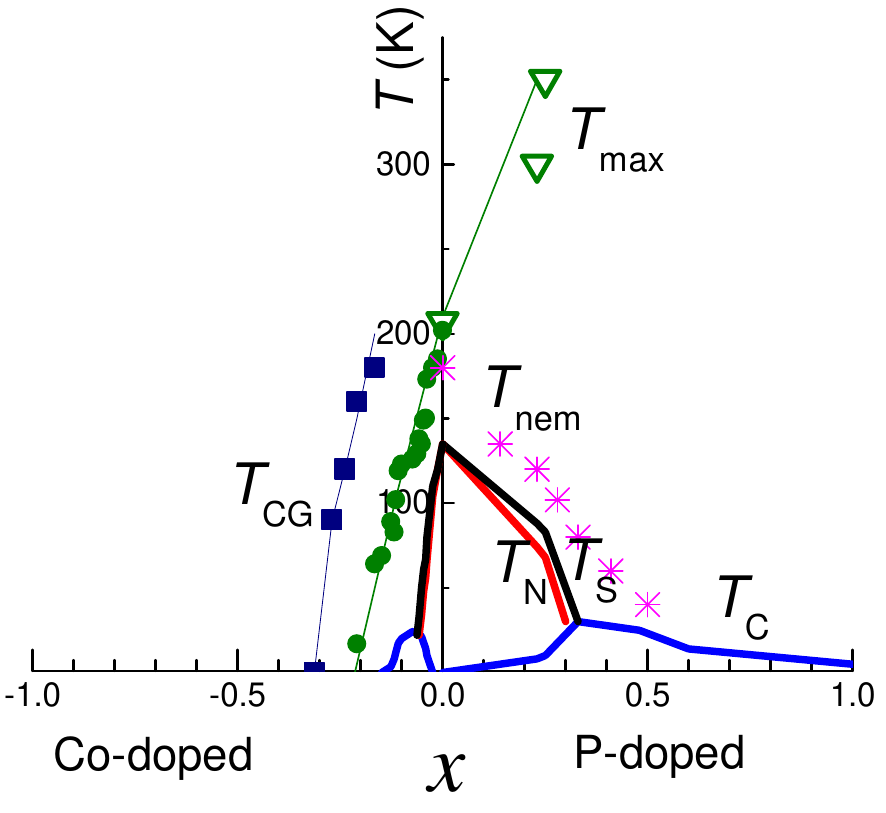}%
\caption{(Color online) Temperature-composition $x$ phase diagram of electron- doped BaFe$_{1-x}$Co$_x$As$_2$ and isoelectron-substituted BaFe$_2$(As$_{1-x}$P$_x$)$_2$ as determined from temperature-dependent inter-plane resistivity measurements. The crossover maximum temperature $T_{max}$ found in the $\rho _c(T)$ of the parent BaFe$_2$As$_2$ (see top curve in Fig.~\ref{interplane}), shifts in a very different way for various types of doping: it is rapidly suppressed with electron-doping (green solid circles) \cite{pseudogap} but rapidly increases  (green down triangles) for isoelectron substitution. An additional minimum feature in $\rho_c(T)$ of heavily overdoped BaFe$_{1-x}$Co$_x$As$_2$ defines a characteristic temperature $T_{CG}$, not found in isoelectron-doped BaFe$_2$(As$_{1-x}$P$_x$)$_2$. Magenta crosses show an onset temperature of nematic anomaly in in-plane resistivity and torque measurements \cite{KasaharaNature}, black and red lines show temperatures of structural tetragonal-to-orthorhombic, $T_S$, and antiferromagnetic, $T_N$, transitions, respectively.
}
\label{phaseD}%
\end{figure}

In Fig.~\ref{phaseD} we summarize a doping evolution of the characteristic temperatures of the $c$-axis resistivity: maximum $T_{max}$, temperatures of the structural, magnetic and superconducting transitions, for electron- (BaCo122) and isoelectron- doped (BaP122) BaFe$_2$As$_2$ compounds. For the latter we also show temperatures of nematic transition found in magnetic torque measurements \cite{KasaharaNature}. The $\rho_c(T)$ maximum shows a dramatic asymmetry in $x$ for electron-doping and isoelectron substitutions. The crossover temperature is rapidly suppressed with doping in BaCo122, it is preceded by metallic temperature dependence at high temperatures above a minimum in $\rho_c(T)$ for heavily doped BaCo122. A close to $T$-linear $\rho_c(T)$ dependence is found at a critical concentration $x$=0.313 \cite{pseudogap}, and normal metallic $\rho_c(T)$, and temperature-independent Pauli susceptibility, $\chi (T)$, and Hall constant are restored for $x>$0.313 \cite{pseudogap,Hall}.

%%%%%%Figure2 phase diagram for Ba122

%%%%%%Figure3 rhoa and rhoc at optimal doping BaCo, BaK, BaP

\begin{figure}[tbh]%
\centering
\includegraphics[width=8cm]{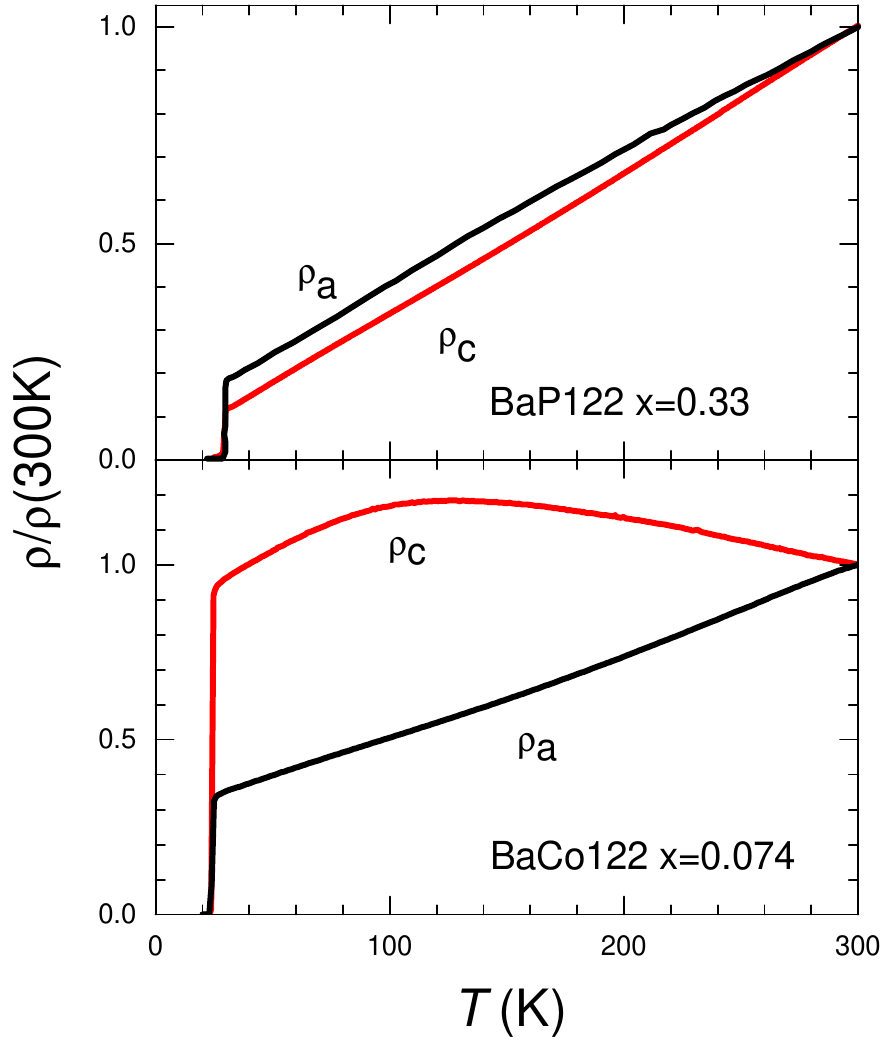}%
\caption{(Color online) Comparison of the temperature-dependent in-plane (data from Ref.~\onlinecite{Kasahara}) and inter-plane resistivity of isoelectron-substituted BaFe$_2$(As$_{1-x}$P$_x$ (top panel) with the e.g. electron- doped BaFe$_{1-x}$Co$_x$As$_2$ (bottom panel) at optimal doping level, with $x$ as indicated. Note that both $\rho _a(T)$ and $\rho _c(T)$ are close to linear just above superconducting $T_c$ for both types of doping, however, the temperature range of the $T$-linear dependence is restricted at higher temperatures by an onset of the a crossover in electron-doped composition, similar to in-plane transport in Ba$_{1-x}$K $_x$Fe$_2$As$_2$, LiFeAs \cite{LiFeAs} and NaFeAs \cite{NaFeAs1,NaFeAs2}. 
}
\label{rhoToptimal}%
\end{figure}

In stark contrast with both these doping dependences, crossover temperature $T_{max}$ shoots up with $x$ of isoelectron P-substitution, and this evolution leads to an interesting difference in the temperature-dependent anisotropic resistivity at optimal doping, as shown in Fig.~\ref{rhoToptimal}. Two panels show $\rho _a(T)$ and $\rho _c(T)$ on a normalized resistivity scale, $\rho(T)/\rho(300K)$, for phosphorus isoelectron-substituted (top panel) and cobalt electron-doped (bottom panel) BaFe$_2$As$_2$. In both cases the resistivity above $T_c$ is close to $T$-linear, but the behavior at higher temperatures differs dramatically, and reveals a clear distinction: the crossover anomalies are absent in the isoelectron-substituted BaP122, while they affect only inter-plane transport in BaCo122.

\section{Discussion}

In BaP122 both in-plane and inter-plane resistivities show the non-Fermi liquid $T$-linear dependence near the optimum doping $x\sim0.3$, a concentration at which the antiferromagnetic quantum critical point has been observed \cite{BaP-QCP}. At higher dopings, $\rho_c(T)$ becomes superlinear at low temperatures and with increasing doping it gradually evolved towards the Fermi-liquid $T^2$ dependence, similar to the doping-evolution found for $\rho_{ab}(T)$ \cite{Kasahara}. The $T$-linear resistivity near the QCP is consistent with the inelastic scattering by two-dimensional (2D) antiferromagnetic fluctuations \cite{Moriya}, which are also evident from the Curie-like temperature dependence of the $1/(T_1T)$ (where $T_1$ is the NMR relaxation time) \cite{BaP-QCP}.

According to the band structure calculations, the Fermi surface of BaP122 comprises five sheets, three hole- and two electron \cite{Kasahara,Currington}. These were observed experimentally by ARPES measurements \cite{ARPES1,ARPES3D} for the whole series of compounds. 
The warping, important for the inter-plane transport, is strongest for the hole- Fermi surface and it increases with 
$x$ in BaFe$_2$(As$_{1-x}$P$_x$)$_2$. This increased warping of the hole sheets has been observed by both ARPES studies \cite{ARPES3D} and quantum oscillations for phosphorus-rich compositions close to $x=1$ \cite{Currington}. The strongest curvature is found near the $Z$ point of the Brillouin zone, for the Fermi surface with dominant contribution of the $d_{z^2}$ orbital, which does not have significant nesting with the electron sheets and thus should be least affected by magnetic fluctuations. The in-plane conductivity is governed by the electron sheets with higher mobility, whereas the inter-plane conductivity is sensitive to the $c$-axis component of Fermi-velocity. 

The observed very similar temperature dependences of the normalized in-plane and inter-plane resistivity with similar residual values (Fig. ~\ref{rhoToptimal}) suggest that the warped $d_{z^2}$ part of the hole sheet also experiences inelastic scattering with non-Fermi liquid $T$-linear temperature dependence, which is counterintuitive considering their position away from hot spots. This observation may suggest that the inter-orbital scattering plays an important role in this system \cite{Shimojima}.

The $T$-linear temperature dependence of $\rho _c(T)$ at optimal doping appears due to a rapid rise of a temperature $T_{max}(x)$ with P-substitution in BaFe$_2$As$_2$, see Figs.~\ref{interplane} and \ref{phaseD}. Although maximum in $\rho_c (T)$  in the layered materials is frequently related to the dimensional crossover \cite{Hussey}, in our study of BaCo122 \cite{pseudogap} we have shown that doping-evolution of the maximum and the appearance of a minimum at high doping levels are inconsistent with this interpretation. The increase with doping of the three-dimensionality of the hole- Fermi-surface does not change the resistivity anisotropy beyond a factor of approximately two uncertainty of the geometric factors with phosphorus substitution from $x$=0 to $x$=0.6. Within this uncertainty, our resistivity anisotropy, $\gamma_{\rho} \approx$6$\pm$2, is in semi-quantitative agreement with the anisotropy of the upper critical field, $\gamma_H$=1.44 in $ac$ magnetization measurements \cite{Gho} and $\gamma_H=$2.49 as found in specific heat study \cite{Welp}, projecting to $\gamma_{\rho}=\gamma_H^2$ of about 2.1 and 6.2, respectively. The anisotropy of the upper critical field $\gamma _H$ shows mere 10 \% change with $x$ variation from 0.3 to 0.55 \cite{Welp}.

In our study of the inter-plane resistivity in transition metal doped Ba122 \cite{pseudogap,pseudogap2} we observed a clear correlation of the maximum in $\rho _c(T)$ with onset of significant temperature dependence of the Knight shift in NMR measurements \cite{Co-pseudogap1,Co-pseudogap2}. Interestingly, this correlation extends to BaP122 as well, with the Knight shift being constant at the optimal doping \cite{BaP-QCP}, and no maximum being observed in $\rho_c (T)$, Fig.~\ref{interplane}. 

Infrared reflectivity measurements from the conducting $ab$-plane find pseudogap features at low temperatures in both optimally doped BaCo122 and BaP122 \cite{SJMoon}. In both cases the feature appears on cooling in the temperature range between 100 and 200 K. While this temperature in BaCo122 is in reasonable agreement with the position of the $T_{max}$ crossover in $\rho _c(T)$, in BaP122 at optimal doping this maximum either moves to above 400~K or is suppressed completely. This fact may be suggestive that either the feature associated with the pseudogap in $ab$-plane spectroscopic reflectivity studies in BaP122 \cite{SJMoon} is related with another anomaly in the normal state of iron pnictides, nematic response above $T_s$ \cite{Fisher,detwinning,KasaharaNature,BaPdetwinning}, or the pseudogap feature observed in the $ab$-plane reflectivity measurements in BaP122 does not affect transport along the $c$-axis . 
Further studies including Hall effect and optical conductivity measurements along the $c-$ axis would be desirable to understand this point. 

\textit{Comparison with the cuprates.} The onset of the pseudogap feature in the cuprates is frequently determined from a temperature of deviations from $T-$linear resistivity \cite{Ando}, a dominant anomalous feature of the normal state transport closely linked with superconducting $T_c$ (see Ref.~[\onlinecite{Taillefer}] for review). Similar analysis was recently done in BaP122 \cite{KasaharaNature} and found a good coincidence with nematic feature found in torque measurements and in-plane resistivity anisotropy in strain-detwinned samples \cite{BaPdetwinning}. In the cuprates the nematic order leads to a two-fold symmetry breaking in the plane as well \cite{Ramzy,JCDavis}. It was suggested that nematicity represents an order parameter for the pseudogap state in the cuprates \cite{JCDavis}.

\section{Conclusions}

Measurements of the inter-plane resistivity in BaP122 show that despite similar suppression of the magnetic/structural transitions with electron-doping  and isoelectron- substitution into Ba122, the crossover maximum feature reveals dramatic difference in response between these cases. The presence/absence of the inter-plane resistivity maximum correlates with the presence/absence of the temperature-dependent NMR Knight shift in BaCo122/BaP122. Contrary to BaCo122, the inter-plane resistivity maximum in BaP122 shows no correlation to nematic anomalies of in-plane resistivity.

%%%%%%%%%%%%%%%%%%%%%%%%%%%% ACKNOWLEDGMENTS
\section{Acknowledgements}

Work at the Ames Laboratory was supported by the Department of Energy-Basic Energy Sciences under Contract No. DE-AC02-07CH11358. Work in Japan was supported by KAKENHI from JSPS, Grant-in-Aid for GCOE program ``The Next Generation of Physics, Spun from Universality and Emergence", Grant-in-Aid from MEXT, Japan.

%%%%%%%%%%%%%%%%%%%%%%%%%%%% BIBLIOGRAPHY

\end{document}